\documentclass[aps,pre,floatfix,twocolumn,showpacs,showkeys,amsmath,amssymb,superscriptaddress,10pt]{revtex4-1}

\usepackage{amsthm}
\usepackage{bm}
\usepackage{epsfig}
\usepackage{verbatim}
\usepackage{graphicx}
\usepackage{grffile}	%More than one point in the graphics file name%
\usepackage{xcolor}

\newcommand{\beq}{\begin{equation}}
\newcommand{\eeq}{\end{equation}}
\newcommand{\beqa}{\begin{eqnarray}}
\newcommand{\eeqa}{\end{eqnarray}}
\newcommand{\beqann}{\begin{eqnarray*}}
\newcommand{\eeqann}{\end{eqnarray*}}

\begin{document}

\title{Nematic phase in the J$_1$-J$_2$ square lattice Ising model in an external field}

\author{Alejandra I. Guerrero}
\email{alejandra.i.guerrero@gmail.com}
\affiliation{Departamento de F\'{\i}sica,
Universidade Federal do Rio Grande do Sul\\
CP 15051, 91501-970 Porto Alegre, RS, Brazil}
\author{Daniel A. Stariolo}
%\email{daniel.stariolo@ufrgs.br}
\affiliation{Departamento de F\'{\i}sica,
Universidade Federal do Rio Grande do Sul and
National Institute of Science and Technology for Complex Systems\\
CP 15051, 91501-970 Porto Alegre, RS, Brazil}
\author{No\'e G. Almarza}
%\email{noe@iqfr.csic.es}
\affiliation{Instituto de Qu{\'\i}mica F{\'\i}sica Rocasolano, CSIC, Serrano 119, E-28006 Madrid, Spain}

\date{\today}

\begin{abstract}
The J$_1$-J$_2$ Ising model in the square lattice in the presence of an external field is studied by two
approaches: the Cluster Variation Method (CVM) and Monte Carlo simulations. The use of the CVM in the square
approximation leads to the presence of a new equilibrium phase, not previously reported for this model: an
Ising-nematic phase, which shows orientational order but not positional order, between the known stripes and
disordered phases. Suitable order parameters are defined and the phase diagram of the model is obtained.
Monte Carlo simulations are in qualitative agreement with the CVM results, giving support to the presence
of the new Ising-nematic phase. Phase diagrams in the temperature-external field plane are obtained for 
selected values of the parameter $\kappa=J_2/|J_1|$ which measures the relative strength of the competing
interactions. From the CVM in the square approximation we obtain a line of second order transitions between
the disordered and nematic phases, while the nematic-stripes phase transitions are found to be of first order.
The Monte Carlo results suggest a line of second order nematic-disordered phase transitions in agreement with
the CVM results. Regarding the stripes-nematic transitions, the present Monte Carlo results are not precise enough to reach
definite conclusions about the nature of the transitions.
\end{abstract}

%\keywords{Nematic phase, CVM, Monte Carlo}

\maketitle

\section{Introduction}
\label{Intro}
Competing interactions are common in many natural and artificial systems, like the presence of conflicting ferromagnetic
and antiferromagnetic interactions in frustrated magnetic systems as spin glasses \cite{FiHe1991} and ultrathin magnetic films
\cite{Abanov95,DeMaWh2000,VaStMaPiPoPe2000}, as well as competition 
between an attractive and a repulsive part in the interaction between atoms and molecules of complex fluids 
\cite{ImRe2006,ArWi2007,OReCReBi2010,Ci2011,PeCiAl2013,PeCiAl2014,AlPeCi20142}. Competition between conflicting
interactions is also relevant in mathematical optimization problems, when decisions have to be made where
not all the constraints can be satisfied simultaneously~\cite{MeMo2009}. Frustration, the inability of a system to statisfy all local constraints, 
is a unifying concept in many natural and artificial systems. 
Competing tendencies usually are responsible for complex behavior like slow relaxation
to equilibrium, strong metastability and rough energy landscapes~\cite{Wales2003}. This makes the study of such systems both
very interesting and challenging. One of the characteristic outcomes of the presence of competing interactions
in a system is the emergence of heterogeneous structures as the equilibrium or low energy states, like stripes,
bubbles, clusters, disordered phases and anisotropic behavior. 

One of the simplest models with competing interactions is the J$_1$-J$_2$ Ising model on the square lattice. This
model is defined as a simple extension of the square lattice Ising model, in which besides the nearest-neighbor
(NN) ferromagnetic or attractive interaction J$_1<0$, a next-nearest-neighbor (NNN) antiferromagnetic or repulsive interaction J$_2>0$ is added. The ground state of the
model depends on the relative intensity of the competing interactions $\kappa=J_2/|J_1|$. For $\kappa <1/2$ it is ferromagnetic
and for $\kappa >1/2$ it has a stripe structure of alternating up and down rows of spins. There is no exact solution
for the thermodynamics of the model. At zero external field it has been studied by a variety of techniques like cluster mean field theory,
transfer matrix and Monte Carlo simulations~\cite{Moran93,Moran94,Cirillo99,dosAnjos08,
Kalz09,Kalz11,Jin12,Jin13,Saguia13}, considering both ferromagnetic $J_1<0$ and antiferromagnetic $J_1>0$ nearest-neighbor 
interactions. The nature of the thermal phase transition from
the stripes to a disordered phase for $\kappa > 1/2$ was controversial. In the most recent studies combining Monte Carlo
simulations and a series of analytical techniques it has  been established that the line of phase transitions in the 
temperature versus $\kappa$ plane is first order for $1/2 < \kappa < 0.67$ and is continuous with Ashkin-Teller critical
behavior for $\kappa > 0.67$. The critical exponents change continuously in this regime between the 4-state Potts model
behavior at $\kappa=0.67$ to standard Ising criticality for $\kappa \to \infty$~\cite{Jin12,Jin13}.

In comparison with the zero field case, the model in an external field has received much less attention.
Queiroz~\cite{Queiroz13} and Yin et.al.~\cite{Yin09} studied the case with both NN and NNN interactions of the antiferromagnetic type,
using transfer-matrix methods in conjunction with finite-size scaling and conformal invariance in the first reference
 and large scale Monte Carlo simulations in the second one.
For $\kappa=1$ the ground state is striped at small fields and a row-shifted phase appears for $4 \leq h \leq 8$. 
The latter state consists of alternating ferro and antiferromagnetically ordered rows (or columns), with the 
ferromagnetic ones parallel to the field. In both studies it was observed a reentrance 
in the boundary stripes-paramagnetic upon lowering the temperature at constant field. 
In Ref. \onlinecite{Yin09} it was argued that the reentrant behavior may be due to the appearance of 
row-shifted ($2\times 2$)  clusters that help to sustain striped ($2\times 1$) order at low temperatures, even for moderately large magnetic fields.
The nature of the phase transitions points to a weak universality scenario with exponents departing slightly from the standard Ising values. 

A natural question when dealing with stripe forming systems is the possibility of existence of an intermediate nematic
phase. A nematic phase in this context is characterized by the presence of orientational order but without translational or positional order \cite{CaMiStTa2006,BaMeSt2013,AlPeCi20142}. In this sense there
are broken symmetry phases but with an intermediate
degree of symmetry, higher than the less symmetric stripe phases in which both orientational and positional orders are present.
Nematic phases associated with intermediate stripe-like order are present in many quasi-two-dimensional systems like ultrathin
ferromagnetic films\cite{Abanov95,CaMiStTa2006,NiSt2007} and electronic liquids in which they may be relevant to understand 
high temperature superconductivity \cite{Han2001,BeFrKiTr2009,FradKiv2010}. 
The conditions under which a system can sustain a nematic phase of this kind are still not completely clear. Strong evidence for the
existence of such phases have been found in systems with isotropic competing interactions at different scales, e.g. when a short
range ferromagnetic interaction competes with a long range antiferromagnetic one decaying with a power law of distance, like the
dipolar interaction \cite{BaSt2009,BaRiSt2013}. In this particular case a nematic phase is present but only quasi-long-range nematic order develops in 2D.
This quasi-nematic phase emerges by breaking a continuous $O(2)$ symmetry, similar to the Kosterlitz-Thouless phase transition in the 
2D XY model. In these kind of models, smectic phases are suppressed at finite temperatures due to the strong fluctuations of the order
parameter. When an external magnetic field is applied, competing dipolar interactions give rise to new and interesting phases.
At small fields stripe phases are still present, although the direction aligned with the field is favoured energetically and it gets
wider as the field is risen until an instability leads to a first order phase transition to a bubble phase at a critical external field value. 
At still higher fields there is a second transition from the bubble to an homogeneously magnetized phase, a saturated paramagnet.
Another salient feature of the field-temperature phase diagram of the system with dipolar interactions in a field is a strong reentrant
behavior. This has been observed in beautiful experiments on ultrathin films of Fe/Cu(001)~\cite{SaLiPo2010,SaRaViPe2010} and also in mean field 
approximations~\cite{PoGoSaBiPeVi2010,CaCaBiSt2011,VeStBi2014}. As for the presence of nematic phases in an external field, this is still
an open question in dipolar or electronic systems.
Compared to the behavior of the dipolar frustrated systems, the J$_1$-J$_2$ model in the square lattice stands at the opposite side: it has a very simple stripe
phase, with long range orientational and positional order, absent in models with long range isotropic interactions. The disordered-stripe
phase transition in this model corresponds to the breaking of the $Z_4$ symmetry of the square lattice to
the $Z_2$ symmetry of the stripe phase. Then, the question we try to answer in this work is: is it possible for the 
J$_1$-J$_2$ model in the square lattice to sustain an Ising-nematic phase? , i.e. a phase with orientational but not positional
order ? To give even a partial answer to such question is always a considerable challenge. This is because the nematic order
parameter in stripe forming systems amounts to compute correlation functions in different space directions searching for a
breaking of isotropy characteristic of these phases \cite{StBa2010,BaSt2011}. Then, the most common analytical approaches for a one-particle
order parameter, namely mean field theory, fails
at detecting nematic phases and one must go beyond naive MFT to an approximation which allows to compute anisotropic correlations.
In this work we have studied the J$_1$-J$_2$ model both with and without an external field by means of two approaches: the
Cluster Variation Method (CVM) which is a cluster mean field theory, and Monte Carlo simulations. The CVM allows for a systematic
improvement upon naive MFT by considering clusters of particles of increasing size in an exact way in the partition function. To
our knowledge, this is the first time the CVM is used with the specific aim of searching for anisotropic correlations, for which
it is particularly well suited. In fact, the first step beyond the mean field approximation in a lattice is the two-site or pair
approximation, also knwon as Bethe-Peierls approximation~\cite{Bethe1935,Tanaka2002}. This amounts to consider in a exact form all clusters with two sites.
This approximation is known to predict correctly that $d=2$ is the lower critical dimension for the Ising ferromagnet. Nevertheless,
because it does not distinguish any geometric or spatial features in the sum over pairs of sites, it is not able to capture rotation
symmetry-breaking, a distinctive feature of orientational phases. The next degree of approximation in the square lattice is the plaquette or
square approximation, which considers exactly clusters of four sites, i.e. first and second neighbors. We will show that this is enough to
capture the presence of nematic phases in models with competing first and second neighbor interactions.
We did not find evidences of nematic phases in the J$_1$-J$_2$ model at zero external field within the square
approximation in the CVM, but a nematic phase appears when a field is present, both in the CVM approach and in Monte Carlo simulations.
\section{The Cluster Variation Method}
\label{cvm}
We give here a very brief description of the Cluster Variation Method, focusing on quantities that
will be useful in the calculation below. There is a large literature on the technical aspects of the method and the interested
reader can refer to references \cite{Kikuchi51,Sanchez84,An88,Tanaka2002} for more comprehensive 
discussions of the method and its potential
for computing variational approximations to the free energy of different systems. 

Consider the variational free energy
\beq
F_t=Tr\,(\rho_t H)+k_BT\ Tr\,(\rho_t \ln{\rho_t}),
\eeq
where $Tr$ means a trace or a sum over all the relevant degrees of freedom of the Hamiltonian $H$, and
$\rho_t$ is a trial density matrix which satisfies the normalization constraint $Tr\,\rho_t=1$. A
systematic way for obtaining variational approximations to the free energy is to express it in terms of a {\em cumulant 
expansion}. Following the exposition by Tanaka \cite{Tanaka2002}, consider the n-body reduced density matrix
for an $N$ body system:
\beq \label{redcond}
\rho_t^{(n)}(1,2,\ldots ,n)=Tr_{n+1}\,\rho_t^{(n+1)}(1,2,\ldots ,n,n+1)
\eeq
for $n=1,2,\ldots ,N-1$, with the normalization conditions:
\beq \label{rednorm}
Tr\,\rho_t^{(1)}(i)=1, \hspace{2cm} i=1,2,\ldots,N.
\eeq
The {\em cluster functions} $G^{(n)}$ and the  {\em cumulant functions} $g^{(n)}$ are defined as
follows:
\beqa 
G^{(1)}(i)&=&Tr\,\left[\rho_t^{(1)}(i)\ln{\rho_t^{(1)}(i)}\right]=g^{(1)}(i),\\
G^{(2)}(i,j)&=&Tr\,\left[\rho_t^{(2)}(i,j)\ln{\rho_t^{(2)}(i,j)}\right] \nonumber \\
        &=&g^{(1)}(i)+g^{(1)}(j)+g^{(2)}(i,j),\\
G^{(3)}(i,j,k)&=&Tr\,\left[\rho_t^{(3)}(i,j,k)\ln{\rho_t^{(3)}(i,j,k)}\right] \nonumber \\
        &=&g^{(1)}(i)+g^{(1)}(j)+g^{(1)}(k)+g^{(2)}(i,j)\nonumber \\
        &+&g^{(2)}(j,k)+g^{(2)}(i,k)+g^{(3)}(i,j,k),
\eeqa
and so on. The largest cluster fuction $G^{(N)}$ (which corresponds to the N-body entropy
term in the variational free energy) can be written as a sum over all the cumulant functions in the form:
\begin{widetext}
\beqa \label{cumexp}
G^{(N)}(1,\ldots,N)&=&Tr\,\left[\rho_t^{(N)}(1,\ldots,N)\ln{\rho_t^{(N)}(1,\ldots,N)}\right] \nonumber \\
   &=&\sum_i g^{(1)}(i)+\sum_{i< j}g^{(2)}(i,j)+\sum_{i<j<k}g^{(3)}(i,j,k)+\cdots + g^{(N)}(1,\ldots, N).
\eeqa
\end{widetext}

In this way the variational free energy can be written in terms of an expansion in cumulant functions:
\beqa \label{varpot}
F_t &=&Tr\,\left[ H\rho_t^{(N)}(1,2,\ldots ,N)\right]+ \nonumber \\
  & & k_BT\left[\sum_i g^{(1)}(i)+\sum_{i<j}g^{(2)}(i,j)+ \right. \nonumber \\
      & &\left. \sum_{i<j<k}g^{(3)}(i,j,k)+\cdots +
           g^{(N)}(1,\ldots, N)\right].
\eeqa
This form of the variational free energy allows to obtain systematic approximations to the true free
energy of a model system by considering a maximal size of cluster to be summed exactly in the partition
function. This is called the parent cluster. This amounts to truncate the cumulant expansion to a given degree and optimizing the
resultant expression with respect to the reduced density matrices
 which must satisfy the reducibility and 
normalization conditions (\ref{redcond}) and (\ref{rednorm}) respectively. 
The reduced density matrices represent all the subclusters contained within the parent cluster.
In some applications a convenient way of implementing the variational
approximation is to parametrize the density matrices in terms of correlation functions and consider
these as variational parameters instead. 

In the case of a system with Ising spins $\{S_i=\pm 1\}$, the reduced density matrices $\rho_t^{(n)}$ can be written as
\cite{Cirillo99}:
\beq
\rho_t^{(n)}=2^{-n}\left[1+\sum_{k}\sigma_k\zeta_k \right]
\eeq
where the sum runs over all subclusters with $k$ sites within cluster $n$ , $\sigma_k=\prod_{i\in k}S_i$ and the k-point correlation 
functions are defined by $\zeta_k=Tr\, \sigma_k \rho_t^{(k)}$. The variational parameters $\zeta_k$ must satisfy:
\beq
\frac{\partial F_t}{\partial \zeta_k}=0.
\label{state.eq}
\eeq
A hierarchy of approximations to the free energy can be constructed in this way. The simplest one corresponds to the 1-point
approximation for the density matrices, the usual mean field approximation. The 2-point approximation is usually called
Bethe-Peierls approximation~\cite{Bethe1935,Tanaka2002}. As discussed in the Introduction, the pair approximation is not able
to detect orientation-dependent features of the equilibrium phases.
In this work, we implemented the 4-point approximation in the square lattice, which is able
to capture the emergence of anisotropic nearest-neighbor correlations or spontaneous rotational symmetry
breaking.

\section{$J_1$-$J_2$ Ising model in the 4-point (square) approximation}
\label{J1J2}

The $J_1$-$J_2$ Ising model on the square lattice is defined by the Hamiltonian:
\beq
{\cal H} =J_1\sum_{\left<xy\right>} S_x S_y+J_2\sum_{\left<\left<xy\right>\right>}S_xS_y
- h\,\sum_x  S_x,
\label{HJ1J2}
\eeq
where $\{S_x=\pm 1, x=1\ldots N\}$ are $N$ Ising spin variables and $h$ is an external field.
$\left<xy\right>$ denotes a sum over pairs of nearest-neighbors and  $\left<\left<xy\right>\right>$ a sum over pairs of 
next-nearest-neighbors. In this work we consider $J_1<0$ and $J_2>0$ representing ferromagnetic NN and 
antiferromagnetic NNN interactions respectively. The competition ratio is defined by $\kappa=\frac{J_2}{|J_1|}\geq 0$.

\begin{figure}[ht!]
\centering
\includegraphics[scale=1.]{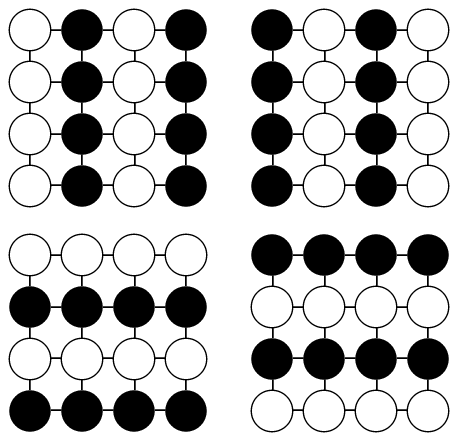}
\caption{Sketch of the ground state configurations for the $J_1$-$J_2$ model without
external field. Empty circles: $S_i=-1$; filled circles:
$S_i=+1$.}
\label{fig.gs}
\end{figure}

At zero external field the ground state of the model is ferro (if $J_1<0$) or antiferromagnetically ordered (if $J_1>0$) 
for $\kappa<1/2$ and striped or superantiferromagnetic if $\kappa>1/2$. In the stripe phase the system can adopt one of four possible
configurations as shown in Figure \ref{fig.gs}.

At zero external field the model has been extensively studied \cite{Moran93,Moran94,Cirillo99,dosAnjos08,
Kalz09,Kalz11,Jin12,Jin13,Saguia13}, considering both ferromagnetic $J_1<0$ and antiferromagnetic $J_1>0$ NN 
interactions and antiferromagnetic $J_2>0$ NNN interactions. The nature of the thermal phase transition from
the stripes to a disordered phase for $\kappa > 1/2$ was controversial. In the most recent studies combining Monte Carlo
simulations and a series of analytical techniques it has  been established that the line of phase transitions in the 
temperature versus $\kappa$ plane is first order for $1/2 < \kappa < 0.67$ and is continuous with Ashkin-Teller critical
behavior for $\kappa > 0.67$. The critical exponents change continuously in this regime between the 4-state Potts model
behavior at $\kappa=0.67$ to standard Ising criticality for $\kappa \to \infty$~ \cite{Jin12,Jin13}.
In the square approximation, the CVM free energy corresponding to the Hamiltonian (\ref{HJ1J2}) is given by 
\cite{Cirillo99}:
\begin{widetext}
\beqa
 F &=&J_1\sum_{\left<xy\right>} Tr \left(S_x S_y \rho_{\left<xy\right>} \right)
+J_2\sum_{\left<\left<xy\right>\right>}  Tr \left(S_x S_y \rho_{\left<\left<xy\right>\right>}\right) 
- h\,\sum_x Tr \left(S_x \rho_x\right)+k_BT \left[ \sum_x g^{(1)}(x)\right. \nonumber\\
&+& \left.\sum_{\left<xy\right>} g^{(2)}(x,y) 
+\sum_{\left<\left<xy\right>\right>} g^{(2)}(x,y)
+\sum_{[xyz]} g^{(3)}(x,y,z)+\sum_{^x_y\Box^w_z} g^{(4)}(x,y,z,w)\right].
\label{cvmJ1J2}
\eeqa
\end{widetext}
In the last equation, the sums over $x$, $\left<xy\right>$, $\left<\left<xy\right>\right>$, $[xyz]$, 
${^x_y\Box^w_z}$ denote sums over all sites, NN pairs, NNN pairs, 
clusters of three sites and squares respectively.
  
Expressing the cumulant functions $g$ in terms of the cluster
functions $G$ the variational free energy reads:
\beqa
 F &=&J_1\sum_{\left<xy\right>} Tr \left(S_x S_y \rho_{\left<xy\right>} \right)
+J_2\sum_{\left<\left<xy\right>\right>}  Tr \left(S_x S_y \rho_{\left<\left<xy\right>\right>}\right) \nonumber \\
&-& h\, \sum_x Tr \left(S_x \rho_x\right)+ k_B T\left[ \sum_x G^{(1)}(x) \right. \nonumber\\
&-&\left. \sum_{\left<xy\right>} G^{(2)}(x,y) +\sum_{^x_y\Box^w_z} G^{(4)}(x,y,z,w) \right].
\label{FJ1J2g}
\eeqa

Following the general approach outlined in Section \ref{cvm} it is convenient to write the variational free energy in terms of correlation functions given by:
\beqa
m_x&=&Tr\left( S_x\rho_x\right) \nonumber\\
l_{xy}&=&Tr \left(S_x S_y \rho_{\left<xy\right>} \right)\nonumber\\
c_{xy}&=&Tr \left(S_x S_y \rho_{\left<\left<xy\right>\right>}\right) \nonumber\\
k_{yxw}&=&Tr \left(S_y S_x S_w\rho_{\left[yxw\right]} \right)\nonumber\\
d_{xyzw}&=&Tr \left(S_x S_y S_z S_w \rho_{^x_y\Box^w_z}\right),
\eeqa
which are related to the reduced density matrices by:
\beqann
\rho_x&=&\frac{1}{2}\left( 1+m_xS_x\right) \nonumber\\
\rho_{\left<xy\right>}&=&\frac{1}{4}\left(1+m_xS_x+m_yS_y+l_{xy}S_xS_y\right)\nonumber\\
\rho_{\left<\left<xy\right>\right>}&=&\frac{1}{4}\left(1+m_xS_x+m_yS_y+c_{xy}S_xS_y\right)\nonumber\\
\eeqann
\beqa
\rho_{^x_y\Box^w_z}&=&\frac{1}{16}(1+m_xS_x+m_yS_y+m_zS_z+m_wS_w\nonumber\\
&+&l_{xw}S_xS_w+l_{wz}S_wS_z+l_{zy}S_zS_y+l_{xy}S_xS_y\nonumber\\
&+&c_{xz}S_xS_z+c_{yw}S_yS_w+k_{yxw}S_yS_xS_w \nonumber\\
&+&k_{xwz}S_xS_wS_z+k_{wzy}S_wS_zS_y +k_{zyx}S_zS_yS_x\nonumber\\
&+&d_{xyzw}S_xS_yS_zS_w).
\eeqa

Substituting these definitions onto (\ref{FJ1J2g}) we get the variational free energy of the $J_1$-$J_2$ model in the CVM 
square approximation\cite{Cirillo99}:
\beqa
 F&=&J_1\sum_{\left<xy\right>} l_{xy} 
+J_2\sum_{\left<\left<xy\right>\right>} c_{xy} - h\,\sum_x m_x\nonumber\\
&+&k_B T \left[ \sum_x Tr \left(\rho_x log \rho_x\right) 
- \sum_{\left<xy\right>}Tr \left(\rho_{\left<xy\right>} log \rho_{\left<xy\right>}\right) \right. \nonumber\\
&+&\left.\sum_{^x_y\Box^w_z} Tr \left(\rho_{^x_y\Box^w_z} log\rho_{^x_y\Box^w_z}\right) \right].
\label{FJ1J2}
\eeqa

After computing the traces one is left with an expression for the variational free energy in terms of a set
of correlation functions representative of the approximation considered. The form of equation (\ref{FJ1J2})
makes clear that up to the pair approximation the two directions in the square lattice enter in a completely
symmetric way, the different pairs of sites are decoupled. It is in the last term, when square plaquettes are
considered, that the coupling between different directions in space can lead to novel behavior. 
The minimization of the free
energy is in general a difficult task. The state (stationarity) equations are given by (\ref{state.eq}).
It is possible to compute the derivatives and try to solve the set of coupled nonlinear equations of state.
Instead of that, we preferred to minimize the variational free energy numerically for given sets
of external parameters.  $k_B=1$ was set for all calculations.

\section{CVM results in an external field}

For $\kappa>\frac{1}{2}$ and small magnetic fields the ground state is striped ($2\times 1$). 
When $J_1$ is ferromagnetic and $J_2$ antiferromagnetic, the stripe order is eventually destroyed by the presence of an external field,
and all the spins become aligned with the field at $h_c=\pm 2(J_1+2J_2)$.
In the case both interactions are antiferromagnetic, for $ -4J_2 \le h \le 4J_2 $ the ground state is still ($2\times 1$), while in the interval 
$4J_2 \le h \le 4J_1+4J_2$ and $-(4J_1+4J_2) \le h \le -4J_2 $ it becomes row shifted ($2\times 2$)\cite{Yin09,Queiroz13}. The latter 
state consists of alternating ferro and antiferromagnetically ordered rows (or columns), with the 
ferromagnetic ones parallel to the field. For higher fields the equilibrium state of the system corresponds to a saturated paramagnet.

\begin{figure*}[ht!]
\centering
\includegraphics[scale=0.4]{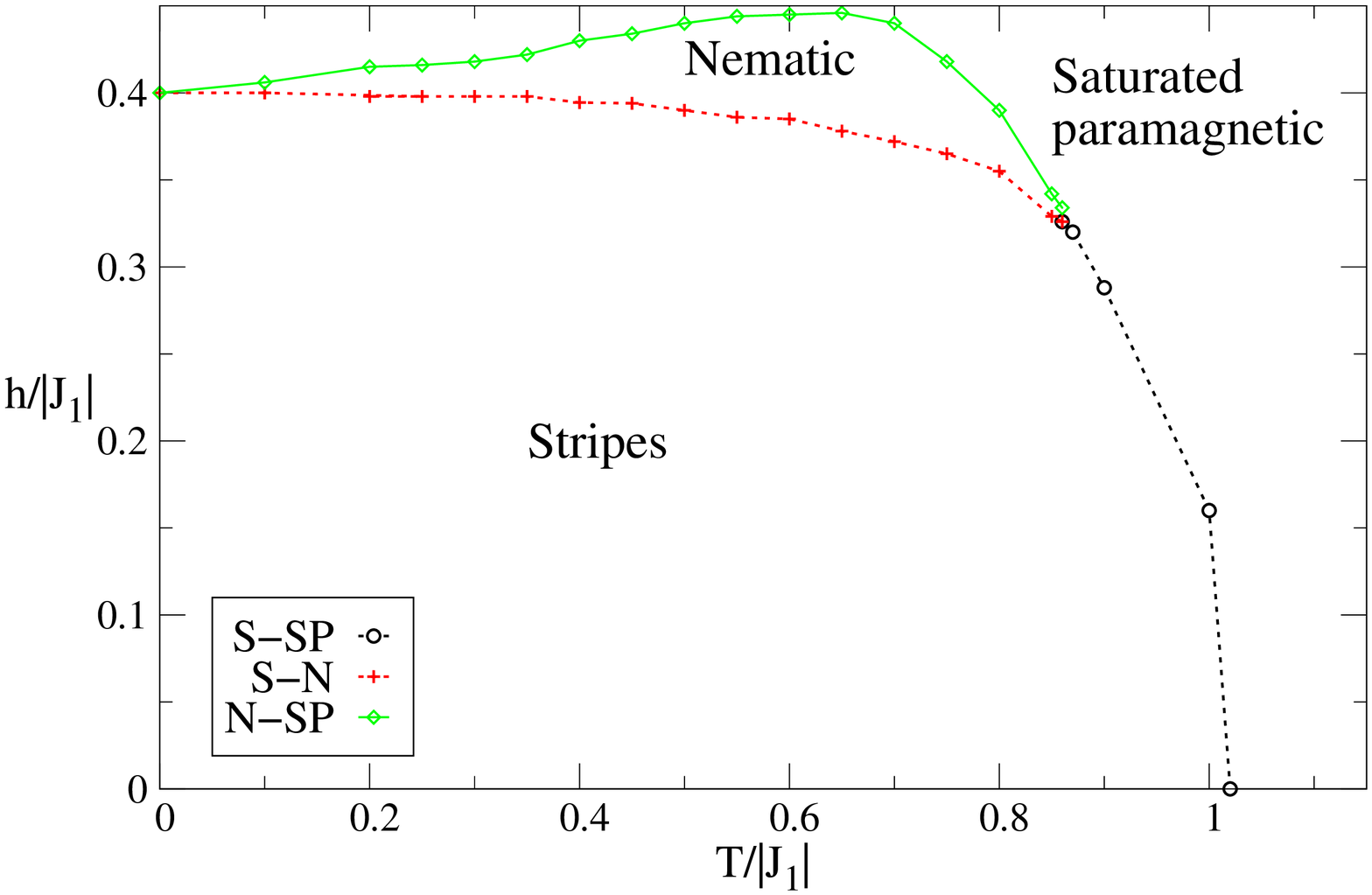}
\caption{(Color online) Reduced external field versus reduced temperature phase diagram for $\kappa=0.6$. Full (dotted) lines correspond to continuous (discontinuous) transitions.}
\label{phasediagk06}
\vspace{1cm}
\includegraphics[scale=0.5]{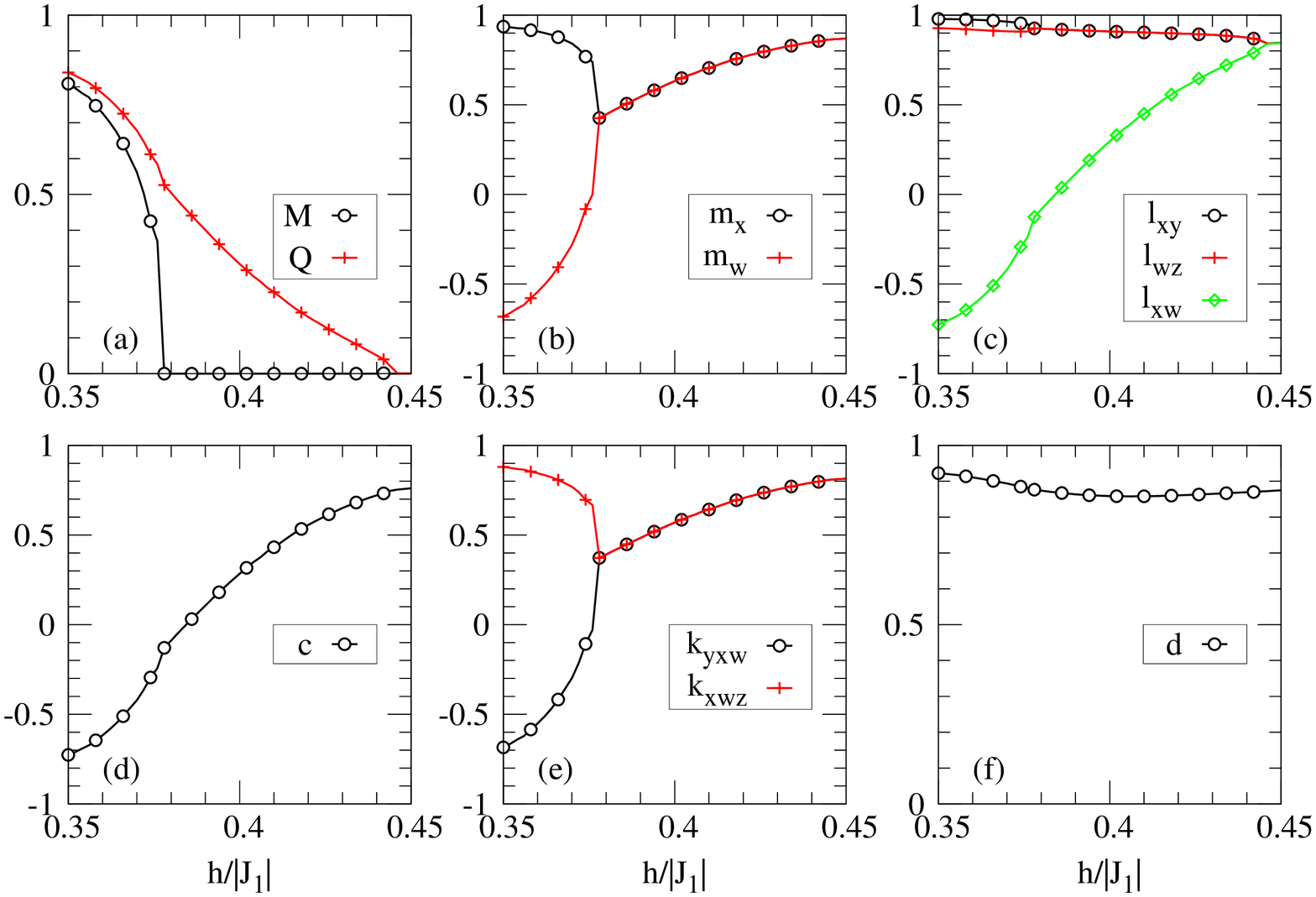}
\caption{(Color online) Correlation
functions versus reduced magnetic field for $\kappa=0.6$ and $T/|J_1|=0.65$. Panel a: positional ($M_s$) and orientational order parameters ($Q$). Panel b: local magnetizations. Panel c: nearest-neighbor correlations. Panel d: next-nearest-neighbor correlations. Panel e: three site correlations. Panel f: square correlations.}
\label{CorrHT=0.65}
\end{figure*}

Here we consider the system at finite $h$ with ferromagnetic NN ($J_1<0$) and antiferromagnetic NNN
($J_2>0$) interactions for two different competition ratios $\kappa=1$ and $\kappa=0.6$ for which the ground state is striped. 
With the aim of searching for purely orientational nematic-like phases, i.e. phases without positional order, 
we minimized the CVM free energy of Eq. (\ref{FJ1J2}) for the parameters (related to the elementary square defined
in (\ref{cvmJ1J2})): $m_x=m_y$, $m_w=m_z$, $l_{xw}=l_{yz}$,
$l_{xy}$, $l_{wz}$, $c$, $k_{yxw}=k_{zyx}$, $k_{xwz}=k_{wzy}$ and $d$. This choice implies possible orientational order along
the $xy$ or vertical direction. Note that local magnetizations on horizontal NN sites are allowed to be
different in sign and also in absolute value. Correspondingly, the NN correlation functions $l_{rs}$ may
be different not only between the horizontal and vertical directions but also between the two vertical ones. 
With these choices the values of NNN correlations $c$ and square correlations $d$ are unique. 
The values of $m_{r}$, $l_{rs}$, $c_{rs}$, $k_{rst}$ and $d$ with $r,s,t=x,y,z,w$ that minimize the 
free energy (\ref{FJ1J2}) were calculated numerically for different reduced temperatures ($T/|J_1|$) and reduced external fields ($h/|J_1|$) using the 
routine NMinimize of the software Mathematica \cite{weisstein}.
In order to distinguish between translational (positional) order and orientational order we defined suitable order parameters;
two positional order parameters:
\begin{equation}
 M_F=\frac{2m_x+2m_w}{4}
\label{ferro.op}
\end{equation}
and
\begin{equation}
 M_S=\frac{2m_x-2m_w}{4},
\label{stripe.op}
\end{equation}
describing ferromagnetic and stripe orders respectively. Note that, in case $m_x$ and $m_w$ are both finite but
with different absolute values, a mixed phase with stripe order on a ferromagnetic background can be possible. In this cases,
we have classified these phases as stripe ones.
The orientational order parameter is defined as:
\begin{equation}
 Q=\frac{1}{4}(l_{xy}+l_{wz}-2l_{xw}).
\label{orientational.op}
\end{equation}
In this case orientational order is finite whenever horizontal NN correlations are different from vertical
ones, i.e. when there is a breaking of $Z_4$ symmetry in the NN correlation functions. With these definitions all order 
parameters take values in the range $(-1,1)$.
\begin{figure*}[ht!]
\centering
\includegraphics[scale=0.4]{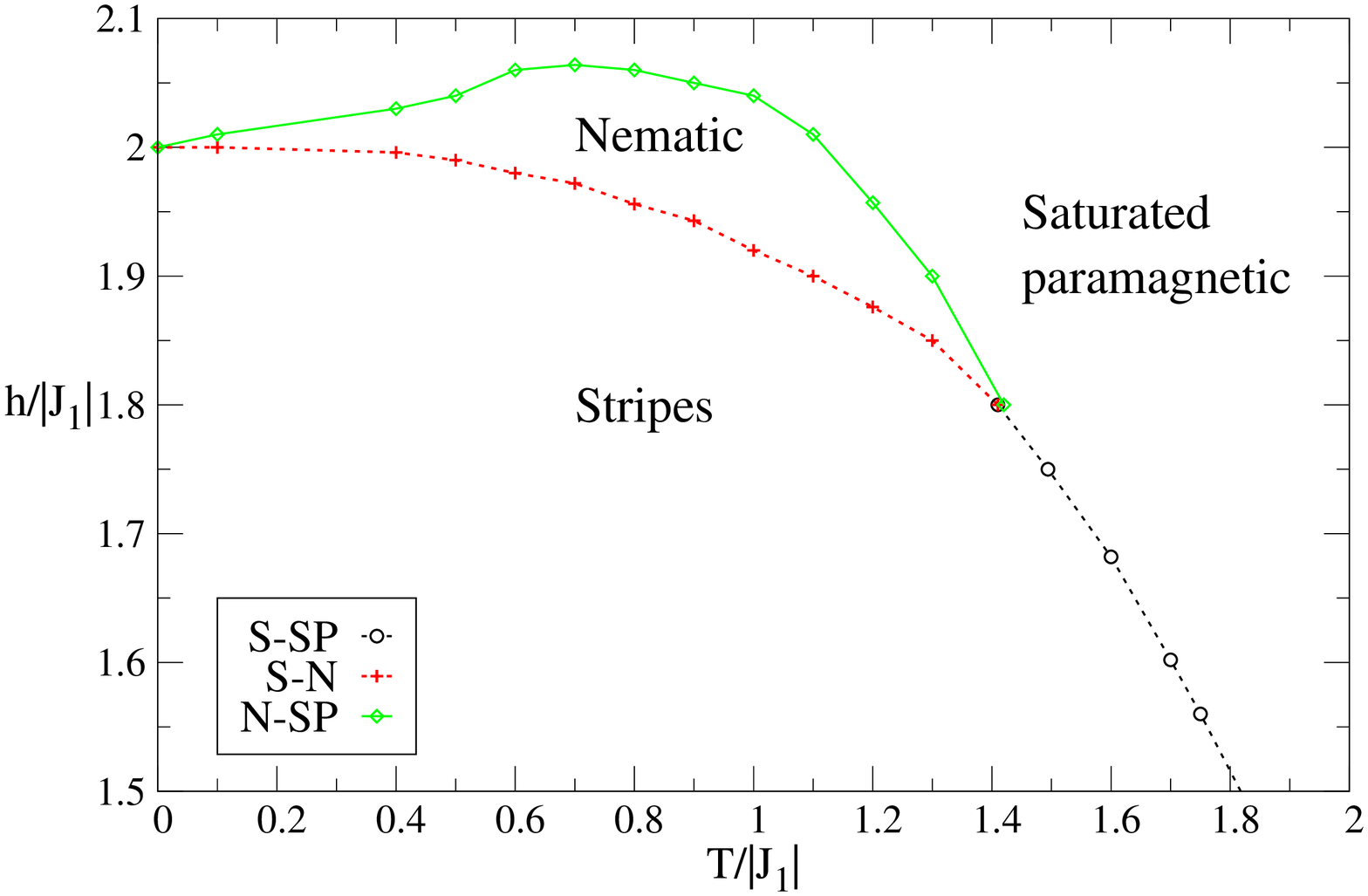}
\caption{(Color online) Reduced external field versus reduced temperature phase diagram for $\kappa=1$. Full (dotted) lines correspond to continuous (discontinuous) transitions.}
\label{phasediagk1}
\vspace{1cm}
\includegraphics[scale=0.5]{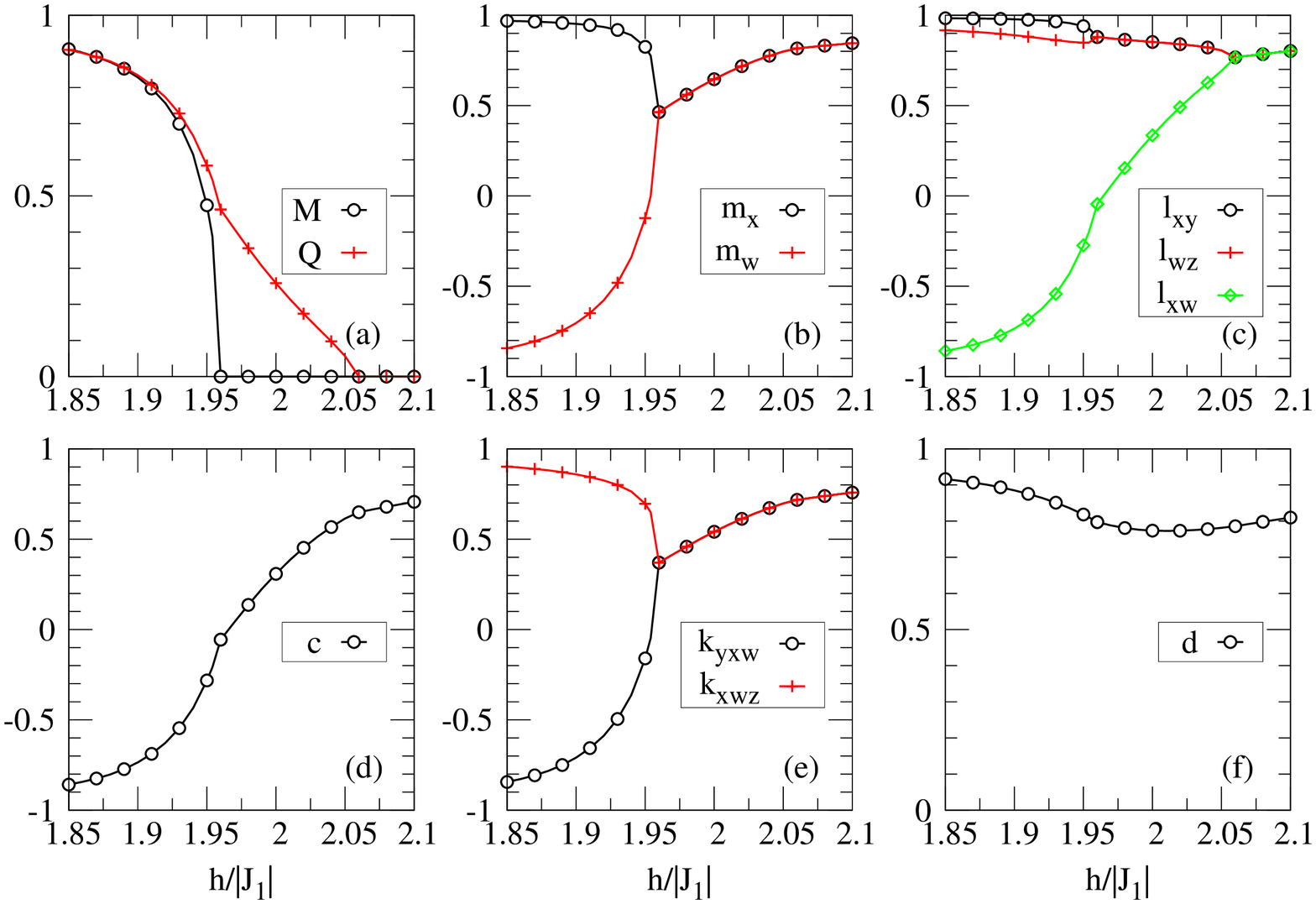}
\caption{(Color online) Correlation
functions versus reduced magnetic field for $\kappa=1$ and $T/|J_1|=0.8$. Panel a: positional ($M_s$) and orientational order parameters ($Q$). Panel b: local magnetizations. Panel c: nearest-neighbor correlations. Panel d: next-nearest-neighbor correlations. Panel e: three site correlations. Panel f: square correlations.}
\label{CorrHT=0.8}
\end{figure*}
The phase diagram in the $h/|J_1| - T/|J_1|$ plane for $\kappa=1$ is shown in figure \ref{phasediagk1}.
For this value of $\kappa$ the ground state is striped. A first difference in this phase
diagram with respect to those in references \onlinecite{Yin09,Queiroz13} is the absence of the row-shifted phase
at large fields. This is due to the ferromagnetic character of the NN interactions in the present work. The absence
of the ($2\times 2$) phase is probably related with the absence of reentrance of the stripe phase in this case, at
variance with the results reported for the model with antiferromagnetic NN interactions~\cite{Yin09,Queiroz13},
as pointed out above. At $h/|J_1|=0$, the 4-point approximation predicts a first order transition line for $\kappa<1$ and 
a second order transition line for $\kappa \geq 1$ \cite{Moran93,Moran94,Cirillo99}.
For $\kappa=1$ and finite $h/|J_1|$  we found a line of first order transitions marked 
by the discotinuity of the order parameter. The discontinuity is observed at both transitions, stripes-saturated paramagnetic 
and stripes-nematic (see Figure \ref{CorrHT=0.8}, first panel). Above the point where the
positional order parameter goes to zero the system still finds itself in a phase with a finite value of the orientational
order parameter $Q$, as seen in Figs. \ref{CorrHT=0.8} and \ref{CorrHT=0.65}. {\em This is the signature of a nematic-like phase
in which the magnetization is homogeneous, unlike in the stripe phase, but correlations show an anisotropic
character, reminiscent of the more ordered stripe phase}. For the case $\kappa=1$ the nematic phase is
observed in the $h/|J_1|-T/|J_1|$ plane in the reduced temperature range $0.1 \leq T/|J_1| \leq 1.4$, as seen in Fig. \ref{phasediagk1}.
The nematic phase terminates in a line of second order phase transitions (green line in Fig. \ref{phasediagk1}) where the system enters a paramagnetic
phase with finite magnetization values due to the external field (saturated paramagnet). The observation
of the nematic phase in the $J_1$-$J_2$ model in an external field is the main result of this work. In the
context of the Cluster Variation Method it is clear that the 4-point approximation is the minimal one which
is able to capture a nematic-like phase of this kind, i.e. a phase with broken orientational symmetry in the correlation
functions. Nevertheless, this possibility was not exploited in previous work within the CVM. 

The behavior of correlation functions for $\kappa=1$ and $T/|J_1|=0.8$ as functions of the external field 
is shown in Figure \ref{CorrHT=0.8}. In the first panel of Figure \ref{CorrHT=0.8} it is seen that the positional and orientational order parameters coincide in the stripe phase, as it should,
but $M_S$ goes to zero before $Q$, signalling a stripe-nematic phase transition at $h/|J_1| \sim 1.96$. The second panel on the upper row
shows that the local magnetizations
$m_x$ and $m_w$ tend to be equal but with opposite signs at very small fields, in agreement with the stripe
character of the ground state for small $h/|J_1|$ values. Nevertheless, they gradually evolve in an asymmetric way until at
the stripe-nematic transition point their values merge in a single one, meaning the onset of an homogeneous
phase with regard to local magnetizations. The anisotropic character of the nematic phase is evidenced
in the third panel on the upper row of Figures \ref{CorrHT=0.8} and \ref{CorrHT=0.65}. It is seen that, within
the stripe phase, correlations along the vertical direction ($l_{xy}$ and $l_{wz}$) are slightly different
reflecting the slightly different values of the local magnetizations at those sites. A change in behavior
between those correlations and the horizontal ones $l_{xw}$ is observed at the stripes-nematic transition
point. Note that if the stripe phase should have ended in a disordered rotationally symmetric state, then
correlations in different directions should merge at this point. This does not happen until a larger
field value where the three different NN correlations considered here merge to a single value at
$h/|J_1| \sim 2.08$. The NNN correlations $c$ display a discontinous derivative at the stripe-nematic transition.

In Figures \ref{phasediagk06} and \ref{CorrHT=0.65} we show the phase diagram and correlation functions 
for $\kappa=0.6$ and $T/|J_1|=0.65$ . The
behavior is qualitatively the same as the case with $\kappa=1$. We show them in order to compare with Monte Carlo
simulation results to be discussed in the next section for $\kappa=0.6$. Although the results from the CVM and Monte Carlo simulations
are qualitatively similar, important quantitative differences arise which should be addressed with higher order
approximations in the CVM approach.

\section{Monte Carlo simulations}
We have performed Monte Carlo simulations of the $J_1$-$J_2$ Ising model on the square lattice to test 
the qualitative consistency of the CVM results. The frustration present in the model turns computer simulations very demanding. 
The simulation procedure combines standard one-site moves\cite{Landau2000} with a cluster method
adapted from the simulation of patchy lattice models \cite{Tavares2014} following the
ideas   of the Wolff algorithm\cite{Wolff1989} in the presence of external fields \cite{Landau2000}.
In the cluster method, one of the spins of the system, and one of the main directions of 
the lattice are chosen at
random. The chosen spin is the starting point (root) of the cluster, then one starts growing the
cluster by adding spins, $j$,  which are NN of the cluster in the chosen direction with probability
 $b=1-\exp(-2|J_1|/k_BT)$ provided that $S_j = S_0$, with $S_0$ being the state of
the spins in the cluster. Once the construction of the cluster is finished acceptance criteria
are applied by taking into account the interaction of the cluster with the remaining spins
of the lattice and the external field\cite{Tavares2014,Landau2000}.
The introduction  of the cluster technique improves appreciably the numerical performance 
of the simulations, specially at low temperature.
In order to enhance further the simulation peformance we have made use of parallel tempering
(or replica exchange) Monte Carlo sampling\cite{Swendsen1986,Earl2005}. This was carried out
as follows: for a given fixed value of the external field (or temperature) we located via
preliminary simulations the approximate value(s) of the temperarature
 (or field) where the transition(s) take(s) place; then we chose an appropriate set of values
of the temperature (field) around such preliminary estimates to carry out the replica-exchange
Monte Carlo simulations. The initial configurations were built by choosing the value of each spin
at random. In order to guarantee the reliability of the final results we run shorter complementary simulations, using ground state configurations as starting point,  to check that the simulations runs
were properly equilibrated.

In the same spirit as with the CVM approach, our main interest was to search for possible nematic phases,
i.e. phases with orientational order but lacking translational order. In order to distinguish 
the presence of such phases we defined suitable order parameters analogous to those defined previously in
the CVM approach.
The translational order parameter (TOP) was defined as the one used in Ref. \onlinecite{Jin13}, in which the configurations
of the system are basically compared with the ground state configurations at zero field.
We analyzed the existence of periodicity in the lattice directions $\hat \alpha=\hat{x},\hat{y}$ by computing the quantities:
\begin{equation}
O_t(\alpha) = \frac{1}{L^2}  \sum_{i=1}^N \left[  2 \times \mod(\alpha_i,2) -1 \right] S_i ;
\end{equation}
where $L$ is the linear size of the square lattice and 
$\alpha_i=x_i, y_i$ are the coordinates of site $i$.
%and $\langle \cdots \rangle$ means a statistical average over equilibrated configurations. 
The global translational order parameter $O_T$ is then defined through:
\begin{equation}
O_T^2 = O_t^2(x) + O_t^2(y).
\end{equation}
Notice that $O_T=1$ for the ground state structures shown in Fig. \ref{fig.gs}. This
order parameter is equivalent to the definition used in (\ref{stripe.op}) which was
suitable for the elementary square of the CVM. 
The translational order can be tested by carrying out a finite-size scaling analysis.
If no translational order exists one expects that
the average of this order parameter $\langle O_T \rangle$  
 will approach zero in
the thermodynamic limit, whereas for the ordered case $\langle O_T \rangle$ will be
finite as $L \rightarrow \infty$. Moreover, if we analyse the translational order
through an isotherm (at varying external field), or as a function of the
temperature (at constant external field), we could expect that the possible
order-disorder transitions involving translational ordering will appear as 
abrupt changes in $\langle O_T\rangle $, specially for large systems.

According to the ground states at zero field (Fig. \ref{fig.gs}), at low
temperatures the system  shows the tendency to form long sequences of spins in the same state
along the main directions of the lattice $\hat{x}$ and/or $\hat{y}$.
The length of these sequences is expected to grow on decreasing the temperature
 and at some point the competition between sequences in the two directions
might lead to a phase transition, in such a way that one of the directions will
be preferred in the ordered phase. In this case the system breaks the fourfold
 (Z$_4$) symmetry of the square lattice reducing it to twofold (Z$_2$) symmetry. This
transition from a disordered Z$_4$ to a Z$_2$ ordered phase is an orientational phase transition and
need not be accompanied with the growth of translational or positional order. So we must distinguish positional
and orientational order parameters, as discussed in relation to the CVM results. 

We can define orientational order parameters (OOP) along directions $\hat{x}$ and $\hat{y}$ as:
\begin{equation}
O_o(\hat{\alpha}) = \frac{1}{L^2} \sum_{i=1}^{L^2}  S( {\bf r}_i ) S({\bf r}_i + \hat{\alpha} );
\hspace{1cm} \hat{\alpha} = \hat{x}, \hat{y};
\end{equation}

We will find that for the GS configurations one of the components of $O_o$ will
have the value $+1$ (that corresponding to the direction in which the sites are
at the same state), whereas the other will take the value $-1$.
A global order parameter is then defined as:
\begin{equation}
O_o = \frac{1}{2} \left[ O_o(\hat{x}) - O_o(\hat{y}) \right].
\end{equation}
This definition is equivalent to Eq. (\ref{orientational.op}) which was suitable in the context
of the square approximation of the CVM. In the thermodynamic limit, finite values of 
the statistical average of its absoute value $<|O_o|>$ 
will indicate that there is a preferential
direction for the sequences of equal spins.

\subsection{Results}

We have considered two values of $\kappa$, namely $\kappa=0.6$, and $\kappa=1.0$.
For $\kappa=0.6$ we have analysed  five cases. In the first one,  $h=0$ was
fixed and we looked at  the variation of different properties with the temperature and the system size. In addition, we took  four values of the reduced temperature $T/|J_1|= 0.80$, 
$0.6667$, $0.50$, and $0.40$, and  looked at the variation of the properties
as a function of the external field, $h$, and the system size at constant temperature.

\begin{figure*}[ht!]
\centering
\includegraphics[scale=0.5]{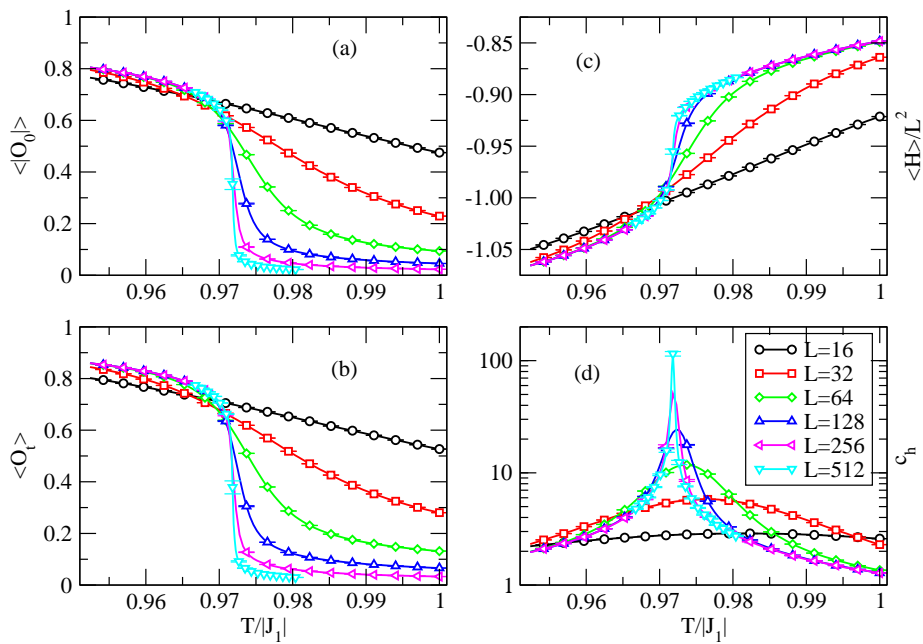}
\caption{(Color online) Results for the $J_1$-$J_2$ model with $\kappa=0.60$ and $h=0$. Different system sizes
(as indicated in the legends) are considered. The results are shown as a function of the reduced temperature. Panel a: Orientational order parameter (OOP); 
Panel b: Translational order parameter (TOP); Panel c: 
mean energy per site $<{\cal H}>/L^2$; Panel d: 
Heat capacity per site at constant field $c_h \equiv (\partial [{\cal H}/L^2] / \partial T)_h$.
}
\label{fig.g060-h0}
\end{figure*}

Some results for $\kappa=0.6$, $h=0$ are shown in Fig. \ref{fig.g060-h0}.
In this case the system exhibits a disordered phase at high temperatures, where
both order parameters tend to zero as the system size increases. At reduced
temperature $T_c/|J_1| \simeq 0.972$ a phase transition occurs.
A direct inspection of the results (See Fig. \ref{fig.g060-h0}) for different properties as a function of $T/|J_1|$ for different
system sizes leads to the following conclusions:
(1)  For $T > T_c$ both order parameters seem to vanish as $L$ grows larger;
whereas for $T < T_c$  both order parameters  converge 
 to a finite value greater than zero for the larger system sizes;
(2) The jump in both order parameters seems to occur at the same value of the temperature ($T_c$);
(3) At $T_c$ there is also a jump in the energy per site: ${\cal H} /L^2$,  (Fig. \ref{fig.g060-h0}.c); 
(4) The corresponding heat capacity at constant field, $ c_h \equiv (\partial [{\cal H}/L^2]/\partial T)_h$
exhibits a clear single peak for systems with  $L\ge 32$, with a value for the maximum that seems to diverge
as $L\rightarrow \infty$ (Fig \ref{fig.g060-h0}.d). 
The scaling of the maximum of $c_h$ for the system sizes considered
does not allow to establish the type of the transition. Two possible scenarios for the
transition are, a 4-state Potts criticality (continuous transitions), with
$c_h \propto L^1$, or a very weak discontinuous transition
as stated by Jin et al.\cite{Jin13}.

\begin{figure*}[ht!]
\centering
\includegraphics[scale=0.5]{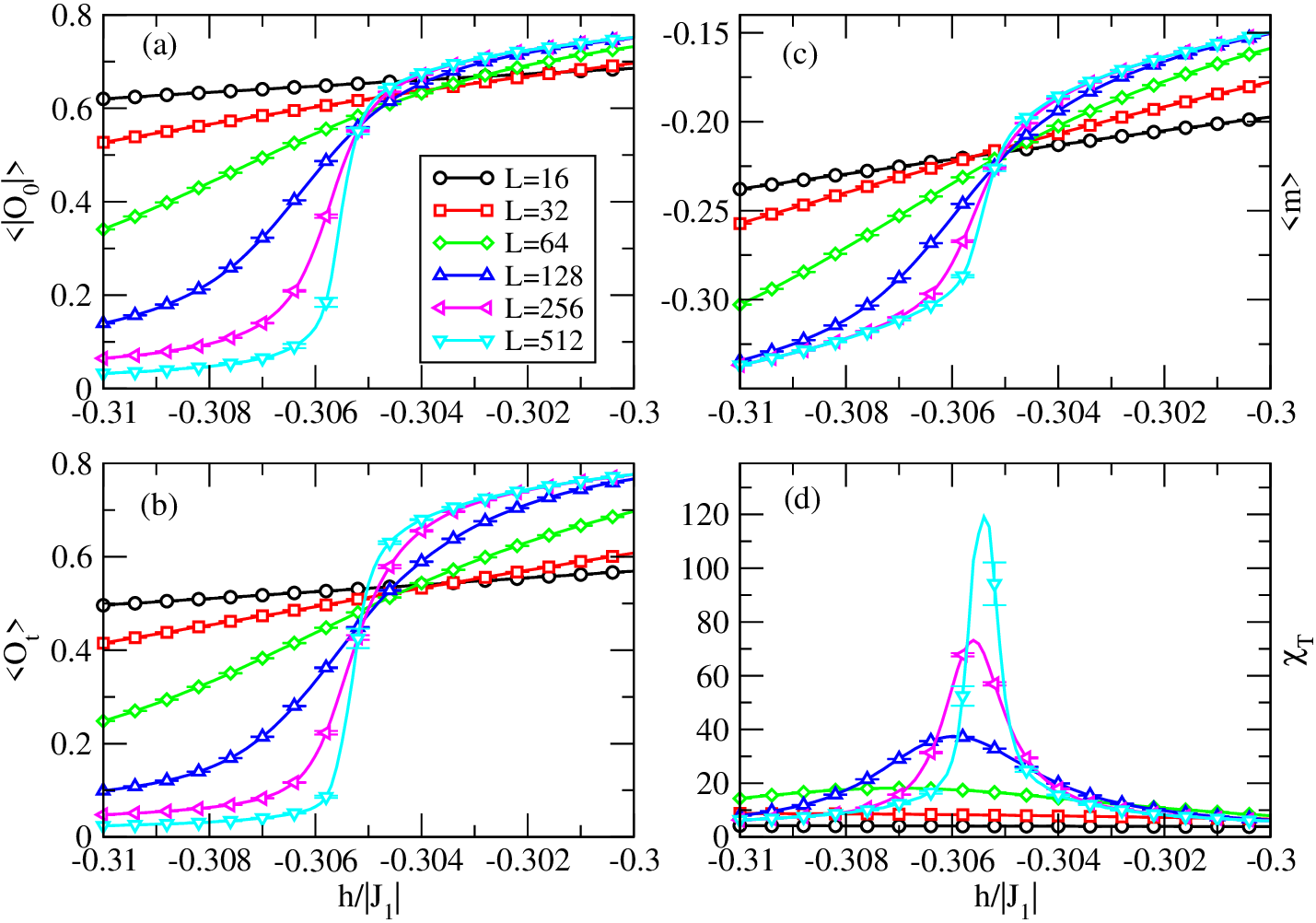}
\caption{(Color online) Results for the $J_1$-$J_2$ model with $\kappa=0.60$ and $T/|J_1|=0.80$. Different system sizes
(as indicated in the legends) are considered. The results are shown as a function of the reduced
external field, $h/|J_1|$. Panel a: Orientational order parameter (OOP); Panel b: Translational order parameter (TOP); Panel c: Magnetization per site, $m$; Panel d: 
$\chi_T \equiv (\partial m / \partial h)_T$.
}
\label{fig.g060-b1.25}
\vspace{1.0cm}
\includegraphics[scale=0.5]{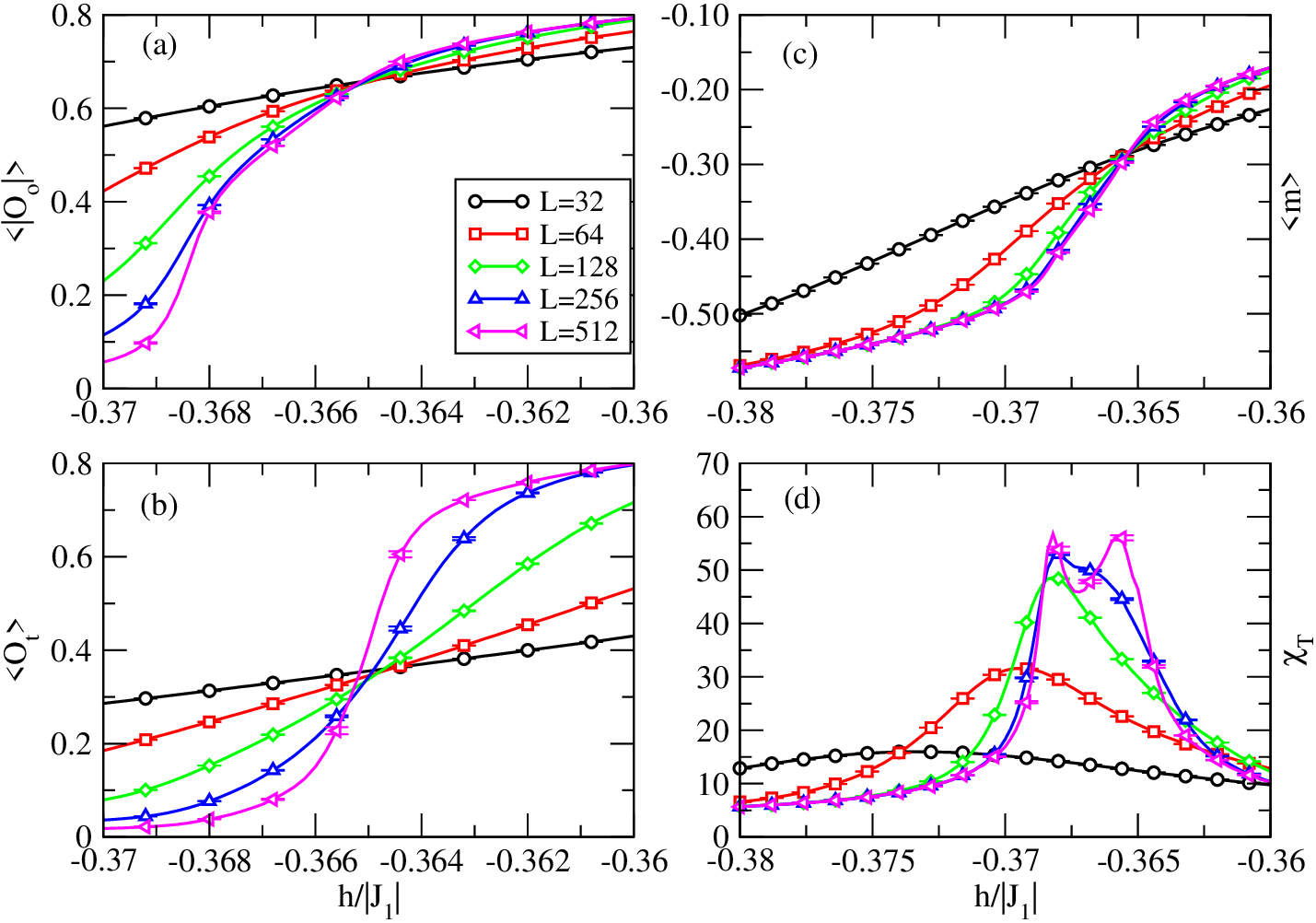}
%\vspace{1cm}
\caption{(Color online) Results for the $J_1$-$J_2$ model with $\kappa=0.60$ and $T/|J_1|=0.6667$.
Different system sizes
(as indicated in the legends) are considered. The results are shown as a function of the reduced
external field, $h/|J_1|$. Panel a: Orientational order parameter (OOP); Panel b: Translational order parameter (TOP); Panel c: Magnetization per site, $m$; Panel d: 
$\chi_T \equiv (\partial m / \partial h)_T$.}
\label{fig.g060-b1.50}
\end{figure*}

Next, we consider the case $\kappa=0.6$, $T/|J_1|=0.8$ and finite $h$.  The main difference with
the case with $h=0$ is that the averaraged magnetization of the system, defined as
$\langle m \rangle = L^{-2} \langle \sum_{i=1}^{L^2} S_i \rangle  $, does not vanish in the thermodynamic limit.
In Figure \ref{fig.g060-b1.25} we show some of the results for this system. From the
results of the order parameters it can be deduced that  at reduced field $h/|J_1| \simeq -0.305$
the system exhibits an order-disorder transition. The results
suggest that the orientational and translational ordering occurs simultaneously.
At the same value of $h$ it is observed a jump in the magnetization, whose derivative
with respect to the external field at constant temperature, $\chi_T$, seems to diverge as  $L \rightarrow \infty$.
The qualitative behavior of the energy per site  with respect to $h$ (not shown), and
its derivative $c_h$ as a function of $h$ is similar to the
behavior for the case $h=0$ when both functions are plotted as functions of $T$.
Therefore, the features of the transition at $T/|J_1|=0.80$ are similar to those found
at zero field. In both cases the orientational and translational orderings seem to occur cooperatively.
This conclusion seems to be consistent with the presence of an unique peak in 
the susceptibility $\chi_{T}$, as shown in Fig. \ref{fig.g060-b1.25}, panel d.

On decreasing the temperature the qualitative features of the transitions change.
We have carried out simulations at three additional values of reduced temperature: $T/|J_1|=$ $0.6667$, 
$0.50$, and $0.40$ (for $\kappa=0.6$).
Some qualitative
differences arise between the phase transitions in these three low-temperature cases and the preceding two cases. 
In Fig. \ref{fig.g060-b1.50} we show the results for the case $T/|J_1|=0.6667$.
First, the aparent common transition for TOP and OOP seems to split into two separated
transitions, i.e. for a given temperature the jumps of TOP and OOP
 occur at different values of the external field $h$. This scenario
of two successive order-disorder transitions instead of just one transition is
consistent with the  incipient splitting of the peak of the susceptibility $\chi_T$
for the larger systems considered. This behavior is in qualitative agreement with the CVM results for
corresponding parameter values observed in Fig. \ref{CorrHT=0.65}.

We have also explored the phase behavior for $\kappa=1.0$. Three cases were considered
(1) Constant field $h=0$; (2) Constant temperature $T/|J_1|=1.0$, and (3)
Constant temperature $T/|J_1|=0.50$. The  phase behavior is similar to that
found for $\kappa=0.60$. At zero field, a single transition is found at $T_c/|J_1| \simeq 2.08$.
For $T/|J_1|=1.0$, the intermediate nematic  phase (with only
orientational order) does not appear, with the ordering transition occurring at $|h|/|J_1| \simeq 1.88$.
However for $T/|J_1|=0.5$  the Ising nematic phase seems to be stable for a very narrow range
of $h$ in the region around  $|h|/|J_1| \simeq 2.00$.
As in the case with $\kappa=0.60$ the splitting of the isotropic-ordered transition into
two transitions is only clearly observed for quite large system sizes $L \simeq 256$, which prevents us from
reaching conclusions about the nature of both transitions.

\subsection{Cumulant analysis of the order parameter distributions}
In the analysis of the phase transitions of model systems it is quite useful to pay attention
to the ratios between the momenta of the order parameter distributions.
Here we considered the ratios $g_4=<O^4>/<O^2>^2$, i.e. $g_{4t}$ and
$g_{4o}$, which are closely related
with the so-called Binder cumulants~\cite{Landau2000}. The analysis of the system size dependence of
these quantities and, in particular,  the crossings of the curves of $g_4$ versus some thermodynamic field 
($T, h, \cdots$) for different system sizes is  often
a very good choice to locate the phase transitions.
\begin{figure*}[ht!]
%\centering
\includegraphics[scale=0.4]{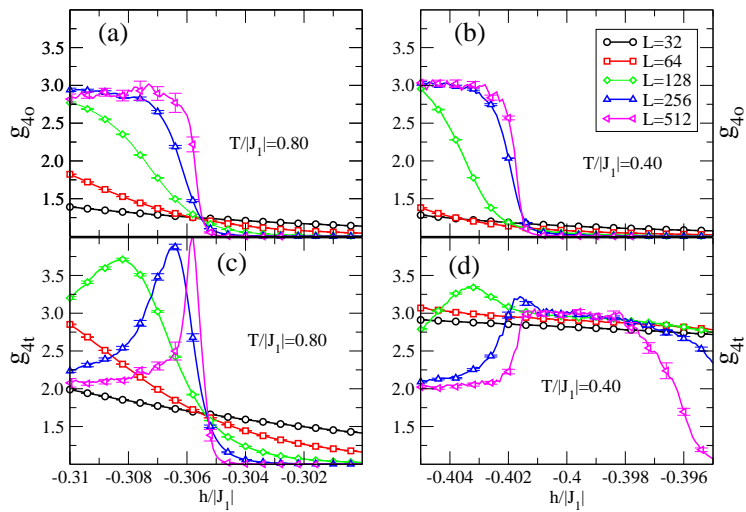}
\caption{(Color online) Order parameter cumulants, $g_4$ as a function of the
reduced external field for two reduced temperatures, different system sizes (see the labels
and legends in the plots) and $\kappa=0.6$.}
\label{fig.g4}
\end{figure*}
In Figure \ref{fig.g4} we show the curves $g_4(h)$ at constant temperatures for
both order parameters and several system sizes. We can appreciate substantial
qualitative differences in the shape of the $g_4(h)$ functions defined on the 
order parameter $O_t$ from the cases $T/|J_1|=0.80$ (one transition),
and $T/|J_1|=0.40$ (two transitions with an intermediate nematic phase) and $\kappa=0.6$.
For $T/|J_1|=0.80$ the transition between the ordered phase ($g_{4t} \approx 1$ )to
the disordered phase $g_{4t} \rightarrow 2 $ occurs quite abruptly as the system size
increases, and the curves for different system sizes seem to cross at $h/|J_1| \simeq -0.305$,
which coincides (or it is quite close) to the crossing point of the  corresponding
lines for the $g_{4o}$ ratio. Notice that the maximum of the susceptibility $\chi_T$
(See Fig. \ref{fig.g060-b1.25}) seems to happen exactly at the same value of $h/|J_1|$.
In the case of $T/|J_1|= 0.40$ the departure of $g_{4t}(h)$ from the ordered phase value,
$g_{4t}=1$, occurs at values of $|h|/|J_1|$ clearly smaller than the corresponding
departure of $g_{4o}(h)$. 
Looking at the cases with $T/|J_1|=0.40$, we can observe that the values of $g_{4o}$ at the crossings
of the curves $g_{4o}(T)$ for different system sizes are consistent with the criticality of the two-dimensional Ising  universality class
\cite{Salas2000}, as one could expect from the symmetry of the order parameter $O_o$.
In addition, the results for the largest systems indicate that
in the range of $h/|J_1|$ values between the two transitions (as predicted by the maxima
of the susceptibility $\chi_T$)  $g_{4t}(h)$ 
 exhibits a plateau,
with values consistent with $g_{4t} \simeq 3 $ in the region where the nematic
phase is supposed to be stable.
The same type of results are found for the cases at $\kappa=1$.

\subsection{Gallery of configurations}

In order to illustrate the differences between the stripe, nematic and disordered phases described in
this work, in what follows we will present some representative configurations of the simulated systems
for $L=256$, considering the lattice gas version of the model. 
The following rules were applied to plot the configurations:
(1) We consider only {\it occupied} sites (or $\sigma_i=1$);
(2) We plot segments between pairs of NN sites if, and only if,  both
sites are occupied; and (3) 
Four colors are considered, depending on the direction of the bond ($\hat{x}$ or $\hat{y}$),
and for each direction depending on the value of the complementary coordinate.
Each color is related with each of the four configurations in the ground state
shown in Fig. \ref{fig.gs}.
%\begin{enumerate}
%\item Bonds in direction $\hat{x}$ over value of coordinate $y$ even.
%\item Bonds in direction $\hat{x}$ over value of coordinate $y$ odd.
%\item Bonds in direction $\hat{y}$ over value of coordinate $x$ even.
%\item Bonds in direction $\hat{y}$ over value of coordinate $x$ odd.
%\end{enumerate}
From this representation of the configurations, we expect for isotropic phases
segments in both directions and four colors with similar probabilities; for nematic
phases most of the segments will be in one of the directions and two colors
will be predominant; whereas for the ordered phase most of the segments will
have the same direction and the same color.
\begin{figure*}[ht!]
\includegraphics[width=4.0cm]{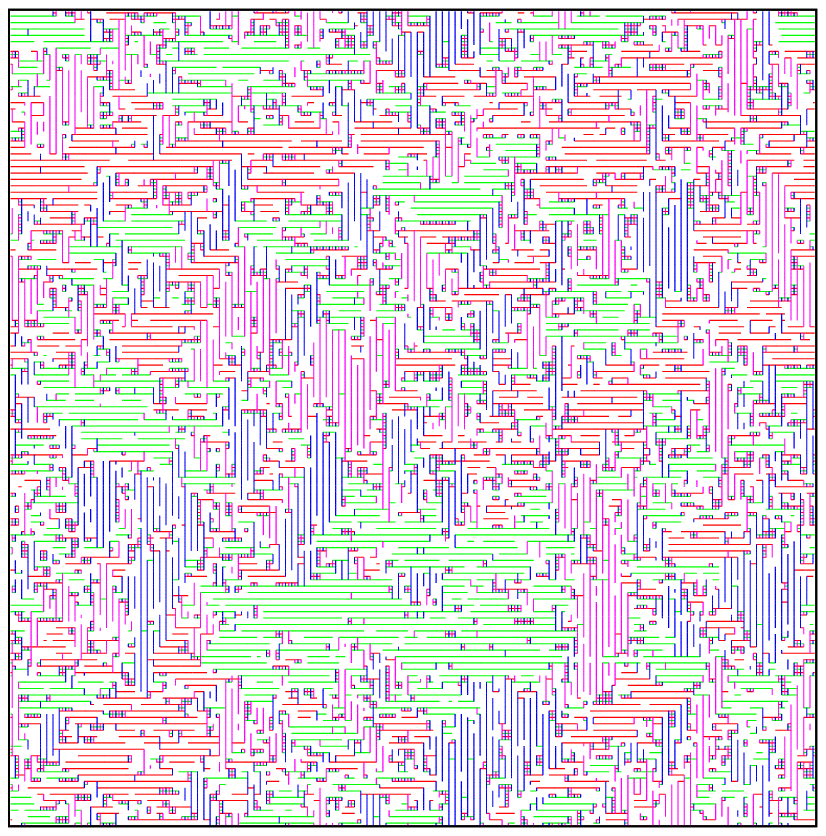}
\includegraphics[width=4.0cm]{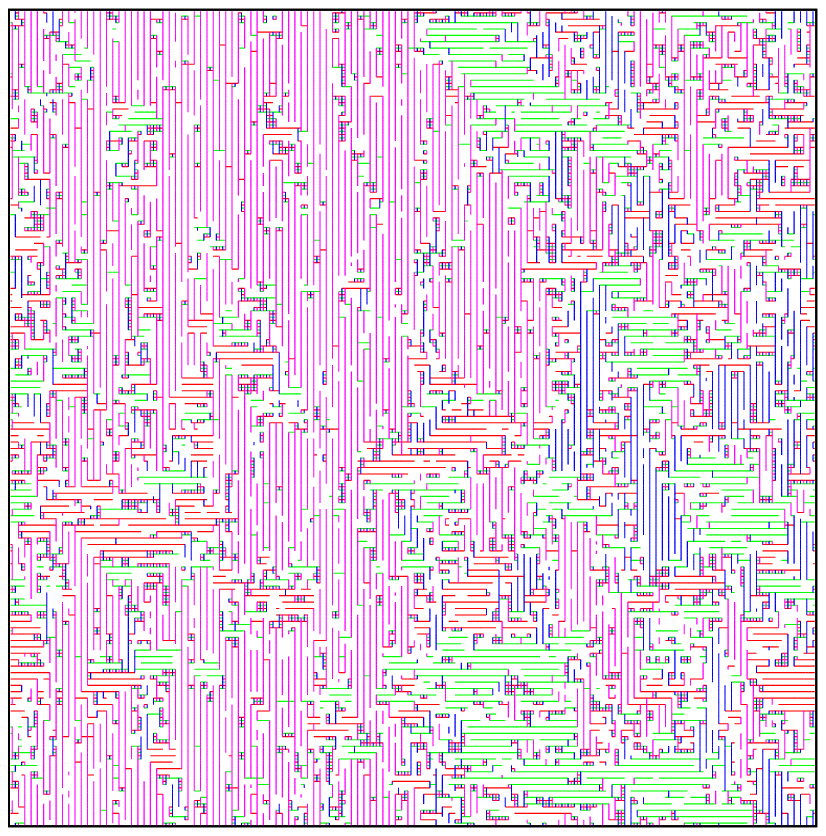}
\includegraphics[width=4.0cm]{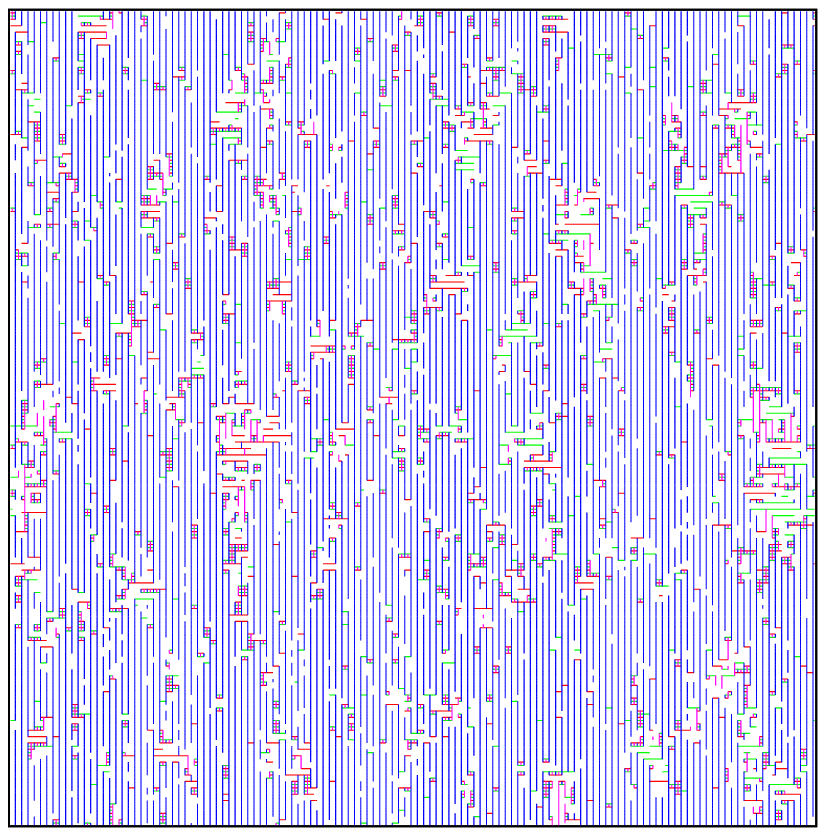}
\caption{(Color online) Representative configurations for $\kappa=0.6$, $h=0$, $L=256$,
close to the order disorder transition. From left to right $T/|J_1|\simeq 0.9756$ (isotropic),
$T/|J_1|=0.9718$ (estimated transition temperature), and $T/|J_1|=0.9662$ (ordered phase
with orientational and translational order).}
\label{fig.confh0}
\end{figure*}
We have chosen two cases. In the first one, shown in FIG. \ref{fig.confh0} we consider fixed value
of the external field, $h=0$  ($\kappa=0.6$) , 
and plot representative configurations for three temperatures in the neighborhood of the transition temperature.
It can be seen that there are no signatures of the presence of the Ising nematic phase.
Above the transition temperature (left panel) one can observe regions in the system where one
of the four colors is predominant. 
At the estimated transition temperature (middle panel in FIG. \ref{fig.confh0}) the system has developed
a large region with most of the bonds in vertical direction and one predominant color, but still relatively
large 
regions with the three remaining colors are still present. As the temperature is further reduced
(right panel in FIG. \ref{fig.confh0}) the regions with minoritary colors appear just as small islands.
\begin{figure*}[ht!]
\includegraphics[width=4.0cm]{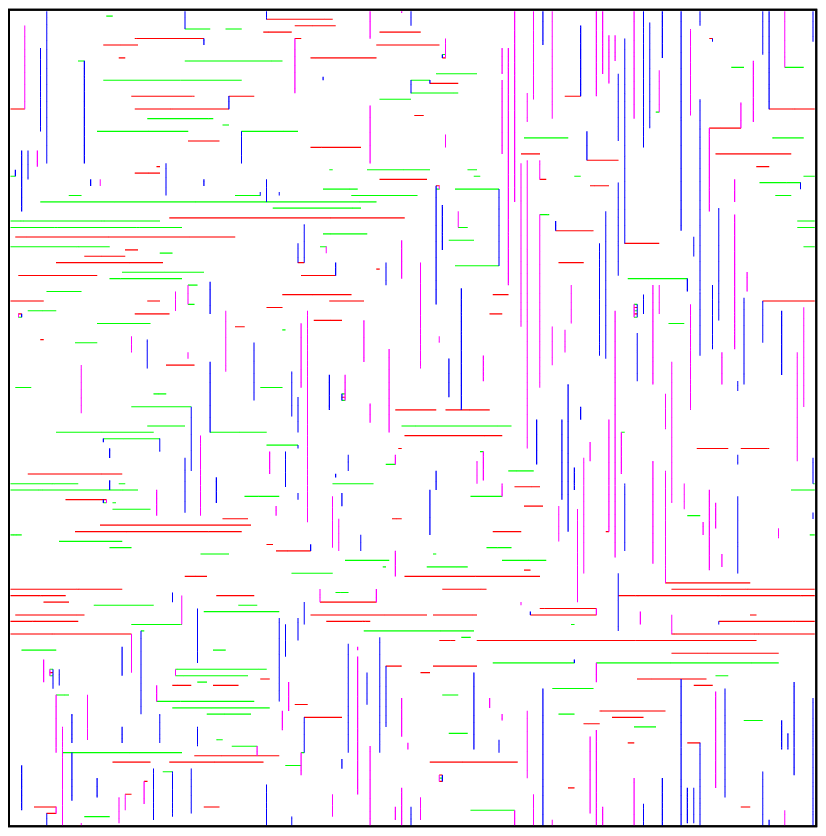}
\includegraphics[width=4.0cm]{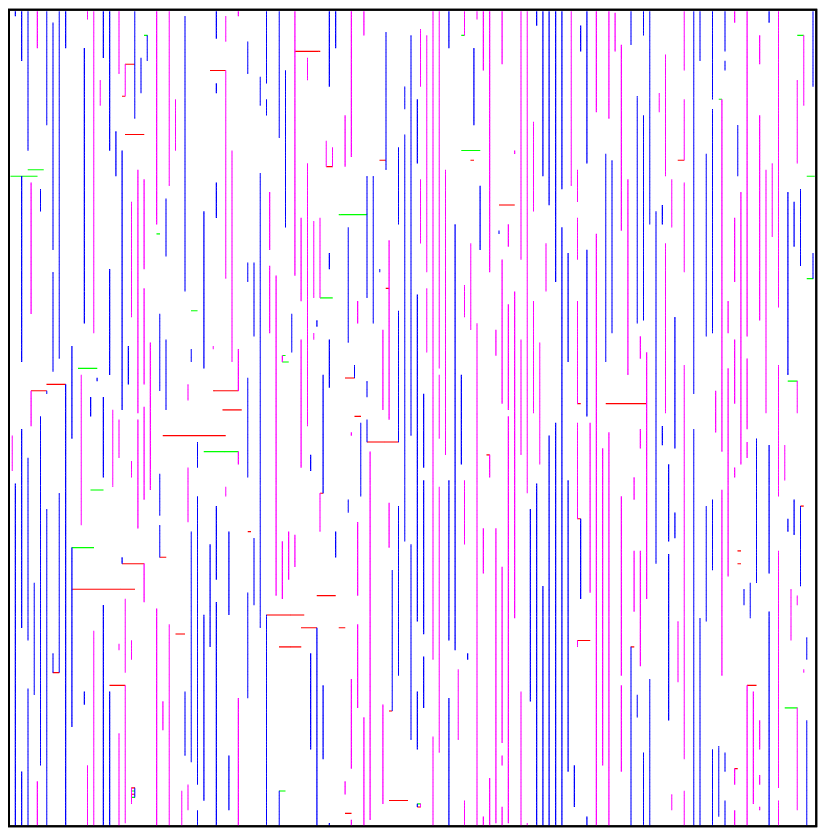}
\includegraphics[width=4.0cm]{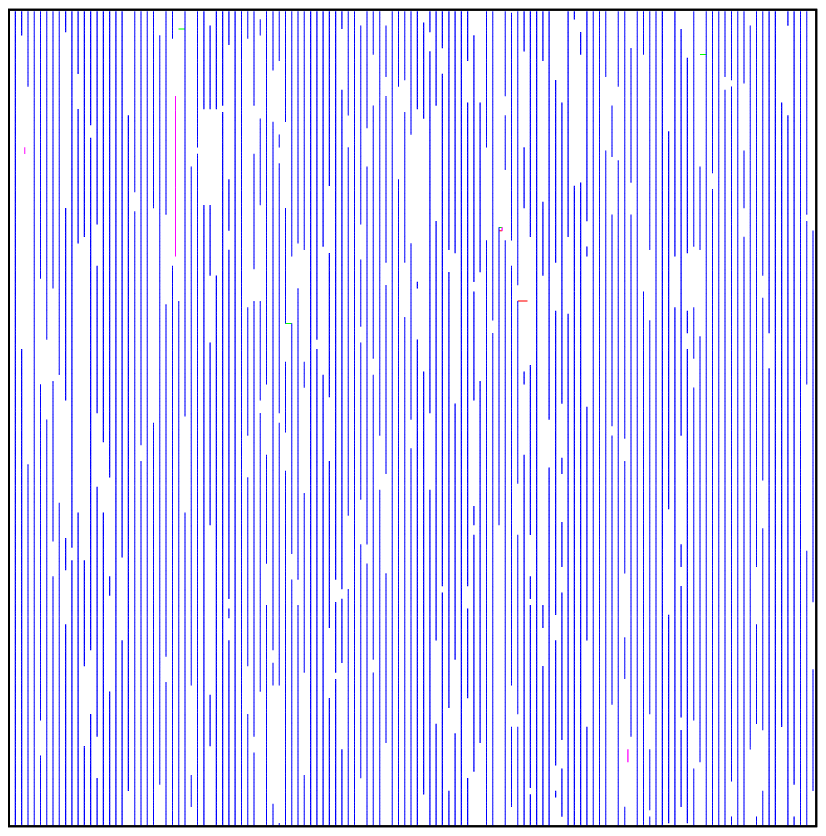}
\caption{(Color online) Representative configurations for $\kappa=0.6$, $T/|J_1|=0.50$, $L=256$,
close to the order disorder transitions. From left to right $h/|J_1|=-0.400$ (isotropic),
$h/|J_1|=-0.397$ (nematic phase), and $h/|J_1|=-0.390$ (ordered phase
with orientational and translational order). The typical length of the rods grows from
left to the right. The densities in the lattice gas model are about (from left to right):
0.12; 0.21; and 0.42.}
\label{fig.confb2}
\end{figure*}
In the second case, shown in FIG. \ref{fig.confb2}, we considered fixed temperature at $T/|J_1|=0.50$ ($\kappa=0.6$).
For this case we expect the stability of an intermediate nematic phase.
Configurations  in the left (isotropic phase) and right (ordered phase) 
panels show similar features, apart from the lower density of segments,
to those found in FIG. \ref{fig.confh0}; however the configuration in the middle panel shows clear
signatures of the nematic phase. Most of the bonds are oriented in vertical direction, but 
none of the two colors associated with this direction is predominant, and no long range order
correlation in horizontal direction can be appreciated.

\section{Conclusions}
\label{conc}
We have shown that the J$_1$-J$_2$ model in an external field has an intermediate phase with only orientational order. 
This is the main result of the present work. We have performed an analytical approach based
on the Cluster Variation Method in the square approximation and compared the results with Monte Carlo simulations.
Both approaches are in qualitative agreement. We did not find  evidence of orientational
intermediate phases of nematic type for zero external field, where our results are compatible with those already
known from the literature. Nevertheless, in the presence of an external field a phase with orientational but without
positional order emerges. This is compatible with an Ising-nematic phase with $Z_2$ symmetry, characterized in this context
by the spontaneous breaking of the $Z_4$ symmetry of the square lattice. 
We found that an Ising-nematic phase exists in a finite window of external field values and temperatures. 

For the parameter values studied we found that the disordered-nematic transition is of second order, with the order
parameter going continuously to zero at $(h_c,T_c)$. Preliminary Monte Carlo results indicate that this transition is
probably in the Ising universality class. The nature of the second transition, from the nematic to a stripe phase
with both orientational and positional orders, is more subtle. The CVM results 
give discontinuous transitions for the parameter values studied. Regarding the Monte Carlo results,
it has been found that there are strong finite size effects. Then, simulations of very large system sizes are required 
to extract definitive conclusions, which are beyond our present capabilities.
The simulation results presented here strongly suggest the existence of a stable
Ising-nematic phase at low temperatures. In principle, according to the form of the order
parameter $O_o$,  and to the
crossings of the $g_{4o}(h)$ functions for different system sizes, one expects a
continuous transition from the disordered to the  nematic phase with 2D Ising criticality.
Regarding the transition between nematic and fully ordered stripe phase 
it seems quite difficult to extract definitive conclusions with the type of calculations 
presented here. There are two basic problems:
first, at intermediate temperatures (say, $T/|J_1|=0.6667$ for $\kappa=0.6$) the 
isotropic-nematic and nematic-stripe phase transitions are very close form
each other, which makes difficult a finite-size scaling treatment.
At lower temperatures, another difficulty arises. The correlation length (the
length of the segments shown in the configuration plots) grows quickly 
as one reduces the temperature. This implies that, again, large systems have
to be considered to extract conclusions.
In addition, the nematic-stripe transition is clearly detected through 
the order parameter $O_t$, the functions $g_{4t}$ or the maxima of the susceptibility
$\chi_T \equiv (\partial m/\partial h)_{T}$ only for large system sizes.

\acknowledgments
A.G.D. and D.A.S. acknowledge partial financial support by Conselho Nacional de Desenvolvimento
Cient\'{\i}fico e Tecnol\'ogico (CNPq), Brazil.
N.G.A. acknowledge the support from the Ministerio de Economía y Competitividad (Spain) under Grant
FIS2013-47350-C5-4-R, and from Acciones bilaterales CSIC-CNPq
(REF: 2011BF0046).
\bibliographystyle{apsrev4-1}
%\bibliography{bibliografPaper,nematics,ultrathin}
\end{document}